\documentclass[a4paper,twocolumn]{article}
\usepackage[a4paper, total={6in, 8in}]{geometry}
\pdfoutput=1
\usepackage[utf8]{inputenc}
\usepackage[english]{babel}
\usepackage[T1]{fontenc}
\usepackage{amsmath}
\usepackage{hyperref}
\usepackage{listings}
\usepackage{tikz}
\usepackage{lipsum}
\usepackage{float}
\usepackage{amsmath,amsfonts,amssymb}
\usepackage{algorithm2e}
\usepackage{algpseudocode}
\RestyleAlgo{ruled}

\usepackage{mathtools}

\DeclarePairedDelimiter\ket{\lvert}{\rangle}
\DeclarePairedDelimiterX\braket[2]{\langle}{\rangle}{#1 \delimsize\vert #2}

\usepackage[
backend=biber,
style=numeric,
sorting=ynt
]{biblatex}
\title{Quantum Algorithm based on Quantum Fourier Transform for Register-by-Constant Addition}

\author{
    \textbf{Filipe Chagas Ferraz} \\
    \textit{Federal University of Mato Grosso (UFMT)} \\
    email: filipe.ferraz@sou.ufmt.br
}

\date{July, 2022}

\begin{document}

\maketitle

\begin{abstract}
    
    Since Shor's \cite{shor} proposition of the method for factoring products of prime numbers using quantum computing, there has been a quest to implement efficient quantum arithmetic algorithms. These algorithms are capable of applying arithmetic operations simultaneously on large sets of values using quantum parallelism. Draper \cite{draper} proposed an addition algorithm based on the quantum Fourier transform whose operands are two quantum registers, which I refer to as register-by-register addition. However, for cases where there is the need to be added a constant value to a target register, Draper's algorithm is more complex than necessary in terms of number of operations (\textit{depth}) and number of qubits used. In this paper, I present a more efficient addition algorithm than Draper's for cases where there needs to be added just a constant to a target register.
\end{abstract}

\section{Introduction}

Quantum computing is an emerging area of technology that addresses techniques for solving computational problems using principles of quantum mechanics. Although there has been a surge of interest and industry activity in quantum computing \cite{preskill2018}, it is not recent that this area has started to be explored. In early 80s, Benioff introduced the quantum version of Turing machines \cite{benioff80, benioff82}. Also in early 80s, the possibility of using quantum machines to simulate quantum systems was addressed by Feynman \cite{feynman82, ladd2010}. In 1985, Deutsch presented a universal quantum computer model capable of simulating finite physical systems and performing fast probabilistic tasks using quantum parallelism \cite{deutsch85}. In the early 90s, Deutsch showed a category of problems that could be solved by quantum computers faster than classical ones \cite{deutsch92}. Other notable pioneer contributions to the field were made by Coppersmith \cite{coppersmith}, Shor \cite{shor}, Vedral et al. \cite{vedral} and Draper \cite{draper}. Coppersmith \cite{coppersmith} proposed the quantum Fourier transform (QFT), an implementation of the discrete Fourier transform for quantum computers, and Shor \cite{shor} proposed a method to factor the product of two prime numbers in polynomial time using quantum computing. The proposition of Shor's method initiated the search for efficient quantum algorithms for performing arithmetic operations such as addition, subtraction, multiplication, and exponentiation with natural numbers stored in quantum registers. The paper by Vedral et al. \cite{vedral} is the first one to propose quantum algorithms for computing the arithmetic operations aforementioned with natural numbers stored in quantum registers using ripple-carry schemes. Later, Draper \cite{draper} proposed a quantum algorithm for summing integers stored in two quantum registers based on the quantum Fourier transform. Draper's \cite{draper} algorithm is a register-by-register adder, that is, both operands of the addition are quantum registers. When it is necessary to add just a constant to the natural values of a target quantum register -- and by ''constant'' I mean an operand that does not have multiple superposed values -- the algorithms of Vedral et al. \cite{vedral} and Draper \cite{draper} are more complex than necessary in terms of the amount of operations (\textit{depth}) and the use of qubits. For this reason, I present in this paper a register-by-constant addition algorithm based on the quantum Fourier transform.

This paper is organized as follows:   

\begin{itemize}
    \item Section 2 briefly adresses the register-by-register addition algorithm proposed by Draper \cite{draper} and explains why this algorithm is more complex than necessary in cases where the intention is to add just a constant to the target register.
    \item Section 3 presents the QFT-based register-by-constant addition algorithm: an addition algorithm based on the quantum Fourier transform (QFT) specific for cases in which there needs to add just a constant to the target register, and less complex than the algorithm proposed by Draper \cite{draper}.
    \item Section 4 presents a mathematical proof of correctness of the register-by-constant addition algorithm presented here.
    \item Section 5 shows an alternative representation of the Draper adder using the register-by-constant addition operator in the Fourier basis.
    \item Finally, section 6 makes some final considerations and concludes the article.
\end{itemize}

\section{Draper's register-by-register addition algorithm}

Draper \cite{draper} proposed a register-by-register addition algorithm based on the quantum Fourier transform. Draper's addition algorithm performs as in the following mapping:

\begin{equation}
    \ket{a,b} \mapsto \ket{a, b + a \pmod{2^N}}
    \label{eq:drapermap}
\end{equation}


In the equation \ref{eq:drapermap}, $a$ and $b$ are natural numbers. $\ket{a}$ and $\ket{b}$ are states each composed by $N$ qubits, i.e., $\ket{a}=\ket{a_N,. ...,a_2,a_1}$ and $\ket{b}=\ket{b_N,...,b_2,b_1}$ such that $a_j \in \{0, 1\}$ and $b_j \in \{0, 1\}$ for any $j \in \{1,...,N\}$.



The capacity for parallel processing of this algorithm is seen when the operand registers have multiple superposed values. For a pair of registers initialized as $\ket{\psi_1} = \sum_{j=0}^{2^N-1} \alpha_j \ket{j}$ and $\ket{\psi_2} = \sum_{k=0}^{2^N-1} \beta_k \ket{k}$, the Draper's adder (represented by $D$) operates as shown in the following equation:

\begin{equation}
    \begin{split}
        D (\ket{\psi_1 } \otimes \ket{\psi_2}) = \\
        D \left[
        \left( \sum_{j=0}^{2^N-1} \alpha_j \ket{j} \right)
        \otimes
        \left( \sum_{k=0}^{2^N-1} \beta_k \ket{k} \right)
        \right] = \\
        D \left( \sum_{j=0}^{2^N-1} \sum_{k=0}^{2^N-1} \alpha_j \beta_k \ket{j, k} \right) = \\
        \sum_{j=0}^{2^N-1} \sum_{k=0}^{2^N-1} \alpha_j \beta_k D \ket{j, k} = \\
        \sum_{j=0}^{2^N-1} \sum_{k=0}^{2^N-1} \alpha_j \beta_k \ket{j, j + k \pmod{2^N}}
    \end{split}
    \label{eq:draperpar}
\end{equation}


Equation \ref{eq:draperpar} shows that when Draper's adder is applied over a pair of registers initialized with multiple superposed values, the result in the second register is a superposition of all possible modular sums $j+k \pmod{2^N}$, each with probability amplitude $\alpha_j \beta_k$.

\begin{figure}[H]
    \centering
    \includegraphics[width=8cm]{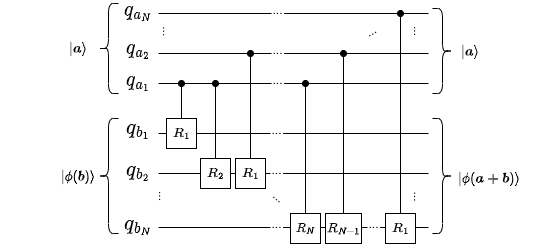}
    \caption{Quantum circuit of the Draper's addition algorithm}
    \label{fig:draper}
\end{figure}

The structure of the Draper's adder consists of a sequence of $\frac{N(N+1)}{2}$ controlled $R_l$ operations between a QFT and an inverse QFT, where $N$ is the number of qubits per operand and $R_l$ is a unitary single-qubit operator defined as:

\begin{equation}
    R_l = \begin{bmatrix}
        1 & 0\\
        0 & e^{\frac{2\pi i}{2^l}}
    \end{bmatrix}
\end{equation}

When there needs to add a constant to a target register of $N$ qubits, i.e., to transform the state $\sum_{k=0}^{2^N-1} w_k \ket{k}$ of the second register into $\sum_{k=0}^{2^N-1} w_k \ket{k + c \pmod{2^N}}$ for $c \in \mathbb{Z}$, Draper's adder requires at least $\lceil\log_2{c}\rceil$ additional qubits for storing the value $c$. In section 3, I present a QFT-based adder with fewer operations and that does not require additional qubits to add a constant to a target register.

\section{Proposed register-by-constant addition algorithm}

\begin{figure}[H]
    \centering
    \includegraphics[width=7cm]{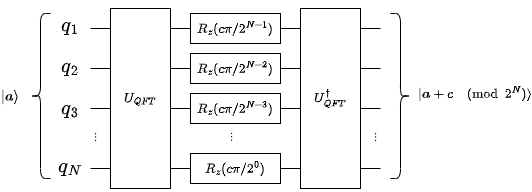}
    \caption{Quantum circuit of the proposed register-by-constant addition algorithm}
    \label{fig:reg-by-const-adder}
\end{figure}

The proposed register-by-constant addition algorithm operates on a single target quantum register, eliminating the need for a second register for storing the constant value. The operation performed by this algorithm is defined for $N$ qubits as the mapping $\ket{a} \mapsto \ket{a + c \pmod{2^N}}$, where $a$ is the natural value stored in the target register before the operation, and $c$ is the constant added to the target register after the operation. A circuit-like representation of this algorithm is presented in figure \ref{fig:reg-by-const-adder}, and a pseudocode implementation is presented in algorithm \ref{alg:reg-by-const-adder}.

Figure \ref{fig:reg-by-const-adder} shows the quantum register-by-constant addition circuit applied over a $N$-qubits register ($q_1, q_2, ..., q_N$) initialized as  $\ket{a}$. In a situation where the register is not initialized with multiple superposed values, $\ket{a}$ can be defined as $\ket{a}=\ket{a_N,...,a_2,a_1}$ such that  $a=\sum_{k=1}^N 2^{k-1} a_k$ and $a_k \in \{0,1\}$. Each qubit $q_k$ stores the respective $a_k$ (note that the register in the circuit in figure \ref{fig:reg-by-const-adder} is big-endian).

\begin{algorithm}
\caption{Implementation in pseudocode of the register-by-constant addition algorithm}\label{alg:reg-by-const-adder}
\KwData{$N \in \mathbb{N}-\{0\}$ \Comment{Number of target qubits}}
\KwData{$c \in \mathbb{Z}$ \Comment{Constant}}
\KwData{$q_1,q_2,...,q_N$ \Comment{Target qubits}}
$U_\text{QFT}(\text{target}=[q_{t}]_{t=1}^N)$ \Comment{Applies a QFT over all target qubits}\;

\For{$t \in [1,...,N]$}{
    $R_z(\theta=\frac{c\pi}{2^{N-t}}, \text{target}=q_{t})$ \Comment{Applies a $R_z(\theta)$ gate over $q_t$} \;
}

$U_\text{QFT}^\dagger(\text{target}=[q_{t}]_{t=1}^N)$ \Comment{Applies a IQFT over all target qubits}\;
\end{algorithm}

The register-by-constant addition algorithm can be represented as the following unitary linear operator:

\begin{equation}
    U_{+}(c) =
    U_{\text{QFT}}^\dagger \times
    U_{\phi(+)}(c)
    \times
    U_{\text{QFT}}
\end{equation}

The operators $U_{\text{QFT}}$ and $U_{\text{QFT}}^\dagger$ are, respectively: the quantum Fourier transform and its inverse; and the operator $U_{\phi(+)}(c)$ is the register-by-constant adder for a constant $c$ in Fourier basis. The operator $U_{\phi(+)}(c)$ is defined for $N$ qubits as:

\begin{equation}
\begin{split}
    U_{\phi(+)}(c) = \bigotimes_{\tau=1}^{N} R_z\left(\frac{c\pi}{2^{N-t}}\right) \\
    = 
    R_z\left(\frac{c\pi}{2^{N-N}}\right) 
    \otimes 
    ...
    \otimes
    R_z\left(\frac{c\pi}{2^{N-1}}\right)
\end{split}
\label{eq:uphiplus1}
\end{equation}

Where $t=N-(\tau-1)$ and $R_z(\theta)$ is a single-qubit \textbf{z}-rotation operator, defined as:

\begin{equation}
    R_z(\theta) = 
    \begin{bmatrix}
        1 & 0 \\
        0 & e^{i\theta}
    \end{bmatrix}
\end{equation}

The application of $U_{+}(c)$ over a register initialized with a state $\ket{\psi}=\sum_{k=0}^{2^N-1} w_k \ket{k}$ is defined as follows:

\begin{equation}
    \begin{split}
        U_{+}(c)\ket{\psi} =
        U_{+}(c) \left( \sum_{k=0}^{2^N-1} w_k \ket{k} \right) \\
        = \sum_{k=0}^{2^N-1} w_k U_{+}(c) \ket{k} \\
        = \sum_{k=0}^{2^N-1} w_k \ket{k + c \pmod{2^N}}
    \end{split}
\end{equation}

The number of operations required by the register-by-constant addition algorithm for $N$ qubits is $T(N)=T_{\text{QFT}}(N) + N + T_{\text{IQFT}}(N)$, where $T_{\text{QFT}}(N)$ is the number of operations required by QFT and $T_{\text{IQFT}}(N)$ is the number of operations required by IQFT. Assuming $T_{\text{QFT}}(N)=T_{\text{IQFT}}(N)=\frac{N(N+1)}{2}$ \cite{barenco_qft}, we have $T(N)=N(N+1)+N=N^2+2N$, and therefore the register-by-constant addition algorithm is $\mathcal{O}(N^2)$.

The Draper's register-by-register addition algorithm with two operand registers of $N$ qubits needs $\frac{N(N+1)}{2}$ controlled operations between the QFT and IQFT, whereas the presented register-by-constant addition algorithm with a target register of $N$ qubits needs only $N$ single-qubit operations between the QFT and IQFT.

\section{Correctness proof of the register-by-constant addition algorithm}

The correctness of the register-by-constant addition algorithm can be demonstrated by analyzing it in matrix form. First, the state of $N$ qubits $\ket{a}$ for $a \in \{x \in \mathbb{N} : 0 \leq x \leq 2^N-1\}$ is expanded in matrix form:

\begin{equation}
    \begin{split}
        \ket{a} 
        = 
        [\lambda_j] 
        \in 
        \mathbb{C}^{2^N \times 1}, 
        \quad
        \lambda_j
        =
        \begin{cases}
            1, \quad j = a \\
            0, \quad j \neq a
        \end{cases}
    \end{split}
    \label{eq:state1}
\end{equation}

In equation \ref{eq:state1}, the expression $[\lambda_j] \in \mathbb{C}^{2^N\times 1}$ corresponds to a matrix of $2^N$ rows and $1$ column. The variable $j$ corresponds to the row index of the respective $\lambda_j$ entry of the matrix, such that $0 \leq j \leq 2^N-1$. 

The operator $U_{\text{QFT}}$, which is the quantum discrete Fourier transform, is expanded in matrix form as:

\begin{equation}
    U_{\text{QFT}} = \frac{1}{\sqrt{2^N}} [\omega^{jk}] \in \mathbb{C}^{2^N\times 2^N}
    \label{eq:qft}
\end{equation}

In the equation \ref{eq:qft}, the expression $\frac{1}{\sqrt{2^N}} [\omega^{jk}] \in \mathbb{C}^{2^N\times 2^N}$ corresponds to a matrix of $2^N$ rows and $2^N$ columns of which entries are $\frac{1}{\sqrt{2^N}}\omega^{jk}$, where $j$ is the row index of the entry and $k$ is the column index of the entry. The symbol $\omega$ is a constant defined as $\omega=e^\frac{2\pi i}{2^N}$ \cite{barenco_qft, nielsen}.

The operation $U_{\text{QFT}}\ket{a}$ results in a column matrix equivalent to the column of index $a$ of the matrix $U_{\text{QFT}}$. Describing this operation algebraically, we have:

\begin{equation}
    U_{\text{QFT}}\ket{a}
    =
    \frac{1}{\sqrt{2^N}} [\omega^{ja}] \in \mathbb{C}^{2^N\times 1}
    \label{eq:qftop}
\end{equation}

In equation \ref{eq:qftop}, the expression $\frac{1}{\sqrt{2^N}} [\omega^{ja}] \in \mathbb{C}^{2^N\times 1}$ corresponds to a column matrix with $2^N$ rows whose entries are $\frac{1}{\sqrt{2^N}}\omega^{ja}$, where $j$ is the row index of the entry. This expression is equivalent to $\frac{1}{\sqrt{2^N}}[..., \omega^{ja}, ...]^\intercal$.

The application of $U_{\text{QFT}}^\dagger$ over $U_{\text{QFT}}\ket{a}$ results in $\ket{a}$ itself, which leads us to the following observation:

\begin{equation}
\begin{split}
    U_{\text{QFT}}^\dagger \left( \frac{1}{\sqrt{2^N}} [..., \omega^{ja}, ...]^\intercal \right)
    = \ket{a}, \\
    0 \leq a \leq 2^N - 1
    \label{eq:iqftop}
\end{split}
\end{equation}

The equation \ref{eq:iqftop} is constrained to $0 \leq a \leq 2^N - 1$. However, for any $a \in \mathbb{N}$, the expression $U_{\text{QFT}}^\dagger \left( \frac{1}{\sqrt{2^N}} [..., \omega^{ja}, ...]^\intercal \right)$ results in $\ket{a \pmod{2^N}}$ and this is due to the periodicity of $a \mapsto \omega^{ja}$. This statement is demonstrated in section 4.1.

Based on equation \ref{eq:iqftop}, it can be deduced that:

\begin{equation}
\begin{split}
    U_{\text{QFT}}^\dagger \left( \frac{1}{\sqrt{2^N}} [...,\omega^{j(a+c)},...]^\intercal \right) = \\
    U_{\text{QFT}}^\dagger \left( \frac{1}{\sqrt{2^N}} [...,\omega^{ja}\omega^{jc},...]^\intercal \right) = \\
    \ket{a+c}, \\
    0 \leq a+c \leq 2^N-1
\label{eq:iqftop2}
\end{split}
\end{equation}

In general, for any $\ket{x}$ such that $0 \leq x \leq 2^N-1$, it is considered that $U_{\text{QFT}}\ket{x} = \ket{\phi(x)}$ and $U_{\text{QFT}}^\dagger \ket{\phi(x)} = \ket{x}$, where $\ket{\phi(x)}$ is a notation for the equivalent of $\ket{x}$ in Fourier basis.


Based on the statement presented in equation \ref{eq:iqftop2}, it comes to the conclusion that, in order to create the mapping $\ket{\phi(a)} \mapsto \ket{\phi(a + c)}$, it is necessary to multiply each entry $\frac{1}{\sqrt{2^N}} \omega^{ja}$ of $\ket{\phi(a)}$ by a factor $\omega^{jc}$. This operation can be described as the matrix product $U_{\phi(+)}(c)\ket{\phi(a)}$, since the operator $U_{\phi(+)}(c)$ is expanded in matrix form as shown in the following equation:

\begin{equation}
\begin{split}
    U_{\phi(+)}(c) = [\lambda_{(j,k)}] \in \mathbb{C}^{2^N\times 2^N}, \\
    \lambda_{(j,k)} = 
    \begin{cases}
        \omega^{jc}, \quad j=k\\
        0, \quad j\neq k
    \end{cases}
\end{split}
\label{eq:uphiplus2}
\end{equation}

The equation \ref{eq:uphiplus2} defines $U_{\phi(+)}(c)$ as a diagonal matrix whose main diagonal entries are $\omega^{jc}$, where $j$ and $k$ are, respectively, the row index and the column index of the entry. This definition can also be written as $U_{\phi(+)}(c)=\text{diag}[1, \omega^{c}, \omega^{2c} ..., \omega^{(2^N-1)c}]$ or simply $U_{\phi(+)}(c)=\text{diag}[...,\omega^{j c},...]$. The equivalence between the definitions of $U_{\phi(+)}(c)$ in equations \ref{eq:uphiplus1} and \ref{eq:uphiplus2} is demonstrated by induction in section 4.2.

Based on the facts presented here, it is possible to conclude that:

\begin{equation}
\begin{split}
    \ket{a + c} = \\
    U_{\text{QFT}}^\dagger \ket{\phi(a + c)} = \\
    U_{\text{QFT}}^\dagger \times U_{\phi(+)}(c) \ket{\phi(a)} = \\
    U_{\text{QFT}}^\dagger \times U_{\phi(+)}(c) \times U_{\text{QFT}} \ket{a} = \\
    U_{+}(c)\ket{a}, \\
    0 \leq a + c \leq 2^N - 1
\end{split}
\end{equation}

\subsection{Modularity}

A function $f(\alpha)=e^{i\alpha}=\cos{\alpha}+i\sin{\alpha}$ has a periodicity of $2\pi$, which means that $e^{i\alpha}=e^{i(\alpha \pmod{2\pi})}=e^{i(\alpha+2\pi n)}$ for any $n\in\mathbb{Z}$. Replacing $\alpha$ with $\theta\frac{2 \pi}{2^N}$, we have $e^{\theta\frac{2 \pi i}{2^N}} = e^{i(\theta\frac{2 \pi}{2^N} + 2 \pi n )} = e^{\theta\frac{2 \pi i}{2^N}}e^{2 \pi n i}$ for any $n \in \mathbb{Z}$.

We have $e^{\theta \frac{2 \pi i}{2^N}}=\omega^{\theta}$ and $e^{2\pi n i} = e^{\left(\frac{2\pi i}{2^N}\right)n 2^N} = \omega^{n 2^N}$, and by assuming that $e^{\theta\frac{2 \pi i}{2^N}} = e^{\theta\frac{2 \pi i}{2^N}}e^{2 \pi n i}$ for any $n \in \mathbb{Z}$, we have $\omega^{\theta} = \omega^{\theta}\omega^{n2^N} = \omega^{\theta + n 2^N}$, therefore $\omega^{\theta} = \omega^{\theta  \pmod{2^N}}$. Based on this, we obtain the following statement from the equation \ref{eq:iqftop}:

\begin{equation}
\begin{split}
    U_{\text{QFT}}^\dagger \ket{\phi(x)} = \\
    U_{\text{QFT}}^\dagger \left( \frac{1}{\sqrt{2^N}} [...,\omega^{jx},...]^\intercal \right) = \\
    U_{\text{QFT}}^\dagger \left( \frac{1}{\sqrt{2^N}} [...,\omega^{j(x \pmod{2^N})},...]^\intercal \right) = \\
    \ket{x \pmod{2^N}}
    \label{eq:iqftopmod}
\end{split}
\end{equation}

Equation 15 implies that for any $0 \leq a \leq 2^N$ and $c \in \mathbb{N}$, we have:

\begin{equation}
    U_+(c) \ket{a} = \ket{a+c \pmod{2^N}}
\end{equation}

\subsection{Demonstration of the equivalence between the definitions \ref{eq:uphiplus1} and \ref{eq:uphiplus2}}

The proposition to be demonstrated is:

\begin{equation}
\begin{split}
    U_{\phi(+)}(c) = 
    \bigotimes_{\tau=1}^{N} R_z\left(\frac{c\pi}{2^{N-t}}\right)\\
    = \text{diag}[...,\omega^{jc},...]
\end{split}
\label{eq:prop}
\end{equation}

Where $t=N-(\tau-1)$, $\text{diag}[...,\omega^{jc},...]$ is a diagonal matrix whose entries are $\omega^{jc}$ and $j$ is the row index of the entry.

The expression $R_z\left(\frac{c\pi}{2^{N-t}}\right)$ can be expanded as:

\begin{equation}
    \begin{split}
    R_z\left(\frac{c\pi}{2^{N-t}}\right) =
    \begin{bmatrix}
        1 & 0 \\
        0 & e^{i\frac{c\pi}{2^{N-t}}}
    \end{bmatrix}
    = \\
    \begin{bmatrix}
        1 & 0 \\
        0 & e^{\left(\frac{2i\pi }{2^N}\right)\frac{c}{2^{1-t}}}
    \end{bmatrix}
    =
    \begin{bmatrix}
        1 & 0 \\
        0 & \omega^{\frac{c}{2^{1-t}}}
    \end{bmatrix}
    \end{split}
\end{equation}

To show that the proposition $\ref{eq:prop}$ is true, a mathematical induction is employed. We begin by expanding $U_{\phi(+)}(c)$ for $N=1$:

\begin{equation}
    \begin{split}
        \bigotimes_{\tau=1}^{N} R_z\left(\frac{c\pi}{2^{N-t}}\right) \bigg\rvert_{N=1}
        = 
        \begin{bmatrix}
            1 & 0 \\
            0 & \omega^{\frac{c}{2^{1-1}}}
        \end{bmatrix}
        \\ = 
        \begin{bmatrix}
            1 & 0 \\
            0 & \omega^{c}
        \end{bmatrix}
    \end{split}
\label{eq:step1}
\end{equation}

The equation \ref{eq:step1} satisfies the proposition \ref{eq:prop}.

Expanding $U_{\phi(+)}(c)$ for $N=2$, we have:

\begin{equation}
    \begin{split}
        \bigotimes_{\tau=1}^{N} R_z\left(\frac{c\pi}{2^{N-t}}\right)
        \bigg\rvert_{N=2}
        = \\
        R_z\left(\frac{c\pi}{2^{N-2}}\right)
        \otimes
        R_z\left(\frac{c\pi}{2^{N-1}}\right)
        = \\
        \begin{bmatrix}
            1 & 0 \\
            0 & \omega^{\frac{c}{2^{1-2}}}
        \end{bmatrix}
        \otimes
        \begin{bmatrix}
            1 & 0 \\
            0 & \omega^{\frac{c}{2^{1-1}}}
        \end{bmatrix}
        = \\ 
        \begin{bmatrix}
            1 & 0 \\
            0 & \omega^{2c}
        \end{bmatrix}
        \otimes
        \begin{bmatrix}
            1 & 0 \\
            0 & \omega^{c}
        \end{bmatrix}
        = \\ 
        \begin{bmatrix}
            1 & 0 & 0 & 0\\
            0 & \omega^{c} & 0 & 0 \\
            0 & 0 & \omega^{2c} & 0 \\
            0 & 0 & 0 & \omega^{c}\omega^{2c} \\
        \end{bmatrix}
        = \\ 
        \begin{bmatrix}
            1 & 0 & 0 & 0\\
            0 & \omega^{c} & 0 & 0 \\
            0 & 0 & \omega^{2c} & 0 \\
            0 & 0 & 0 & \omega^{3c} \\
        \end{bmatrix}
    \end{split}
\label{eq:step2}
\end{equation}

The equation \ref{eq:step2} also satisfies the proposition \ref{eq:prop}.

To proof that the proposition \ref{eq:prop} is true for any number of qubits, an inductive step is conducted on the equation \ref{eq:step3}.

\begin{equation}
\begin{split}
    \bigotimes_{\tau=1}^{M} R_z\left(\frac{c\pi}{2^{M-t}}\right) \bigg\rvert_{M=N+1} = \\
    \begin{bmatrix}
        1 & 0 \\
        0 & \omega^{\frac{c}{2^{1-(N+1)}}}
    \end{bmatrix}
    \otimes
    \text{diag}[...,\omega^{j_n c},...]
    = \\ 
    \begin{bmatrix}
        1 & 0 \\
        0 & \omega^{c2^N}
    \end{bmatrix}
    \otimes
    \text{diag}[...,\omega^{j_n c},...]
    = \\
    \text{diag}[...,\omega^{j_n c},...,\omega^{c2^N}\omega^{j_n c},...]
    = \\
    \text{diag}[...,\omega^{j_n c},...,\omega^{(j_n+2^N)c},...]
    = \\
    \text{diag}[...,\omega^{j_m c},...]
\end{split}
\label{eq:step3}    
\end{equation}

In equation \ref{eq:step3}, it is assumed that the proposition \ref{eq:prop} is true for $N$ qubits and it is verified that, under this assumption, it is also true for $M=N+1$ qubits. The variable $j_n$ corresponds to the row index of a matrix of $N$ rows and columns, and the variable $j_m$ corresponds to the row index of a matrix of $M$ rows and columns.

The equation \ref{eq:prop} is therefore true for any number of qubits.








\section{Rewriting the register-by-register adder using the $U_{\phi(+)}(c)$ operator}


It is possible to rewrite the Draper's register-by-register adder using the $U_{\phi(+)}(c)$ operator defined in sections 3 and 4. This alternative representation is easier to understand than the original one.


As already mentioned in section 2, the register-by-register adder executes the mapping $\ket{a, b} \mapsto \ket{a, b + a \pmod{2^N}}$, where $\ket{a}=\ket{a_N,...,a_1}$ and $a_j \in \{0,1\}$.


As in Draper's adder, this alternative register-by-register adder consists of a series of controlled gates applied on the qubits of the second register between QFT and IQFT, i.e, in the Fourier basis. Since it is possible to define $b+a$ as $b+\sum_{j=1}^{N}a_j 2^{j-1}$, the state of the second register in the Fourier basis can be changed from $\ket{\phi(b)}$ to $\ket{\phi(b+a)}$ by applying a $U_{\phi(+)}(2^{j-1})$ controlled by $a_j$ on this register for each $j\in \{1,...,N\}$.

\begin{figure}[H]
    \centering
    \includegraphics[width=8cm]{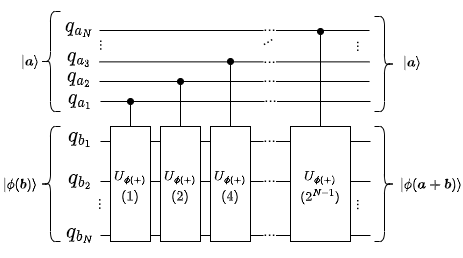}
    \caption{Register-by-register adder using the $U_{\phi(+)}(c)$ operator.}
    \label{fig:rewrittendraperadder}
\end{figure}

Figure \ref{fig:rewrittendraperadder} shows a quantum circuit of the register-by-register adder using controlled $U_{\phi(+)}(c)$ gates.

\section{Conclusion}


This paper presented an arithmetic addition algorithm based on the quantum Fourier transform specific for cases where there needs to be added just a constant to a target register, which is called here register-by-constant adder.  This adder uses fewer qubits than those proposed in \cite{vedral,  draper}, as it requires neither an additional register to store the constant nor ancillary carry qubits. Furthermore, the register-by-constant adder needs fewer operations between the QFT and IQFT than Draper's adder. 


This paper also showed an alternative representation of the Draper's adder using the operator $U_{\phi(+)}(c)$, which is easier to understand than the original one.


For futher works, it might be interesting to benchmark between the register-by-constant adder and other addition algorithms on real quantum hardware to compare the noise of the outputs of these algorithms. Moreover, alternative representations using $U_{\phi(+)}(c)$ of the QFT-based multiplier covered in \cite{RuizPerez17} can be made, and more QFT-based algorithms can be designed using this operator.

\printbibliography

@article{preskill2018,
	doi = {10.22331/q-2018-08-06-79},
	url = {https://doi.org/10.22331%2Fq-2018-08-06-79},
	year = 2018,
	month = {aug},
	publisher = {Verein zur Forderung des Open Access Publizierens in den Quantenwissenschaften},
	volume = {2},
	pages = {79},
	author = {John Preskill},
	title = {Quantum Computing in the {NISQ} era and beyond},
	journal = {Quantum}
}

@article{feynman82,
	doi = {10.1007/bf02650179},
	url = {https://doi.org/10.1007%2Fbf02650179},
	year = 1982,
	month = {jun},
	publisher = {Springer Science and Business Media {LLC}},
	volume = {21},
	number = {6-7},
	pages = {467--488},
	author = {Richard P. Feynman},
	title = {Simulating physics with computers},
	journal = {International Journal of Theoretical Physics}
}

@article{ladd2010,
	doi = {10.1038/nature08812},
	url = {https://doi.org/10.1038%2Fnature08812},
	year = 2010,
	month = {mar},
	publisher = {Springer Science and Business Media {LLC}},
	volume = {464},
	number = {7285},
	pages = {45--53},
	author = {T. D. Ladd and F. Jelezko and R. Laflamme and Y. Nakamura and C. Monroe and J. L. O'Brien},
	title = {Quantum computers},
	journal = {Nature}
}

@article{benioff80,
	doi = {10.1007/bf01011339},
	url = {https://doi.org/10.1007%2Fbf01011339},
	year = 1980,
	month = {may},
	publisher = {Springer Science and Business Media {LLC}},
	volume = {22},
	number = {5},
	pages = {563--591},
	author = {Paul Benioff},
	title = {The computer as a physical system: A microscopic quantum mechanical Hamiltonian model of computers as represented by Turing machines},
	journal = {Journal of Statistical Physics}
}

@article{benioff82,
	doi = {10.1007/bf01342185},
	url = {https://doi.org/10.1007%2Fbf01342185},
	year = 1982,
	month = {nov},
	publisher = {Springer Science and Business Media {LLC}},
	volume = {29},
	number = {3},
	pages = {515--546},
	author = {Paul Benioff},
	title = {Quantum mechanical hamiltonian models of turing machines},
	journal = {Journal of Statistical Physics}
}

@ARTICLE{deutsch85,
    author = {David Deutsch},
    title = {Quantum theory, the Church-Turing principle and the universal quantum computer},
    journal = {Proceedings of the Royal Society of London. Series A: Mathematical and Physical Sciences},
    year = {1985},
    volume = {400},
    pages = {97--117}
}

@article{deutsch92,
  author = {David Deutsch and Richard Jozsa},
  doi = {10.1098/rspa.1992.0167},
  url = {https://doi.org/10.1098/rspa.1992.0167},
  year = {1992},
  month = dec,
  publisher = {The Royal Society},
  volume = {439},
  number = {1907},
  pages = {553--558},
  title = {Rapid solution of problems by quantum computation},
  journal = {Proceedings of the Royal Society of London. Series A: Mathematical and Physical Sciences}
}

@article{vedral,
  title = {Quantum networks for elementary arithmetic operations},
  author = {Vedral, Vlatko and Barenco, Adriano and Ekert, Artur},
  journal = {Phys. Rev. A},
  volume = {54},
  issue = {1},
  pages = {147--153},
  numpages = {0},
  year = {1996},
  month = {Jul},
  publisher = {American Physical Society},
  doi = {10.1103/PhysRevA.54.147},
  url = {https://link.aps.org/doi/10.1103/PhysRevA.54.147}
}

@article{shor,
   title={Polynomial-Time Algorithms for Prime Factorization and Discrete Logarithms on a Quantum Computer},
   volume={26},
   ISSN={1095-7111},
   url={http://dx.doi.org/10.1137/S0097539795293172},
   DOI={10.1137/s0097539795293172},
   number={5},
   journal={SIAM Journal on Computing},
   publisher={Society for Industrial & Applied Mathematics (SIAM)},
   author={Shor, Peter W.},
   year={1997},
   month={Oct},
   pages={1484–1509} }

@misc{coppersmith,
   Author = {D. Coppersmith},
   Title = {An approximate Fourier transform useful in quantum factoring},
   Year = {2002},
   Eprint = {arXiv:quant-ph/0201067},
}

@article{barenco_qft,
	doi = {10.1103/physreva.54.139},
	url = {https://doi.org/10.1103%2Fphysreva.54.139},
	year = 1996,
	month = {jul},
	publisher = {American Physical Society ({APS})},
	volume = {54},
	number = {1},
	pages = {139--146},
	author = {Adriano Barenco and Artur Ekert and Kalle-Antti Suominen and Päivi Törmä},
	title = {Approximate quantum Fourier transform and decoherence},
	journal = {Physical Review A}
}

@misc{draper,
      title={Addition on a Quantum Computer}, 
      author={Thomas G. Draper},
      year={2000},
      eprint={quant-ph/0008033},
      archivePrefix={arXiv},
      primaryClass={quant-ph}
}

@book{nielsen,
  author = {Michael A. Nielsen and Isaac L. Chuang},
  publisher = {Cambridge University Press},
  title = {Quantum Computation and Quantum Information},
  year = 2000
}

@article{RuizPerez17, title={Quantum arithmetic with the quantum Fourier transform}, volume={16}, DOI={10.1007/s11128-017-1603-1}, number={6}, journal={Quantum Information Processing}, author={Ruiz-Perez, Lidia and Garcia-Escartin, Juan Carlos}, year={2017}, month={Apr}}

\end{document}